\documentclass{osa-article}

\journal{oe}

\articletype{Research Article}

\begin{document}

\title{A transportable interrogation laser system with an instability of $\mathrm{mod}\,\sigma_{\rm y} = 3\times10^{-16}$}

\author{Sebastian H\"afner,\authormark{1,2} Sofia Herbers,\authormark{1} Stefan Vogt,\authormark{1,3} Christian Lisdat,\authormark{1} and Uwe Sterr\authormark{1,*}}

\address{\authormark{1}Physikalisch-Technische Bundesanstalt (PTB), Bundesallee 100, 38116 Braunschweig, Germany}
\address{\authormark{2}Currently with Nanosystems and Technologies GmbH, Gleimersh\"auser Str. 10, 98617 Meiningen, Germany}
\address{\authormark{3}Currently with Coherent LaserSystems, Seelandstra{\ss}e 9, 23569 L\"ubeck, Germany}

\email{\authormark{*}uwe.sterr@ptb.de} 



\begin{abstract}
We present an interrogation laser system for a transportable strontium lattice clock operating at 698~nm, which is based on an ultra-low-expansion glass reference cavity.
Transportability is achieved by implementing a rigid, compact, and vibration insensitive mounting of the 12~cm-long reference cavity, sustaining shocks of up to 50~g.
The cavity is mounted at optimized support points that independently constrain all degrees of freedom.
This mounting concept is especially beneficial for cavities with a ratio of length $L$ over diameter $D$ $L/D>1$.
Generally large $L$ helps to reduce thermal noise-induced laser frequency instability while small $D$ leads to small cavity volume.
The frequency instability was evaluated, reaching its thermal noise floor of $\mathrm{mod}\,\sigma_{\rm y} \approx 3 \times 10^{-16}$ for averaging times between 0.5~s and 10~s.
The laser system was successfully operated during several field studies. 
\end{abstract}



\section{Introduction}

Lasers with ultra-high frequency stability are needed as local oscillators for precision measurement instruments like optical clocks.
The clock instability is determined by the interrogation laser through its linewidth \cite{ita93} and through the Dick-effect \cite{dic87}.
A higher clock stability reduces the averaging time, which is highly important to achieve low statistical uncertainties or for time resolved measurements.

Our clock interrogation laser system consists of a laser, whose light is frequency stabilized to an eigenmode of an ultra-stable reference cavity using the Pound-Drever-Hall (PDH) technique.
The high fractional length stability of the thermally and seismically well isolated cavity directly translates to a low fractional frequency instability of the laser light.
The length stability, and therefore the frequency stability of the reference cavity, is fundamentally limited by the thermal noise of its constituents \cite{num04}, with the largest contribution coming from the mirror coatings.
The contribution of the mirrors' displacement noise to the overall thermal fractional length noise of the cavity decreases inversely proportional with the separation of the mirrors. 
Therefore, long cavities allow for a reduced thermal frequency noise floor.
However, they are more difficult to isolate from seismic noise and temperature fluctuations.
Thus far, long cavities have been used more often in well controlled laboratory conditions than in field tests.

The worldwide quest for quieter laser systems has resulted in laser systems based on a cryogenic cooled silicon cavities with a modified Allan deviation down to $\mathrm{mod}\,\sigma_{\rm y} = 4 \times 10^{-17}$ \cite{mat17a, rob19} and a 48~cm long glass cavity with a measured thermal noise floor $\sigma_{\rm y} = 8 \times 10^{-17}$ \cite{hae15a}.
Clock instabilities of below $\sigma_{\rm y}=1\times10^{-16}/\sqrt{\tau/\mathrm{s}}$ \cite{oel19, sch17, alm15} have been realized in a laboratory environment with such high-end clock lasers.

Besides the development of stationary clocks, fundamental science and applications demand transportable clock systems.
Optical clocks with $10^{-18}$ fractional instability linked by optical fibres will enable relativistic geodesy on a centimeter level between two points on Earth \cite{meh18, gro18a}.

Transportable systems need compact, light weight and robust components as well as a rugged design.
Especially for space applications, the clock components have to sustain vibrations of up to 50~g that typically occur during a rocket launch.

To minimize forces induced by accelerations that deform the cavity during operation, in stationary laser systems cavities are mostly placed onto soft support points.
This is not applicable to systems that have to withstand rapid movement without misalignment or even damage.

Transportable cavities need a rigid mounting, which always introduce stress on the mounted spacer. Unavoidable variations of the mounting forces lead to stress and thus to length change of the cavity.
The task is therefore to find a support structure that decouples the cavity length from the constraining forces.
Examples for such mountings are small spherical \cite{lei11, lei13}, cubical \cite{web11}, cylindrical \cite{swi16, dav17, che14a, arg12, vog11} and pyramidal \cite{did18} cavities (25~mm~$<L<$~100~mm) held at optimized points, which have been able to reach low sensitivity to seismic noise while being rigidly mounted to be portable with fractional instabilities of down to $1 \times 10^{-15}$.

In this paper, we report on an interrogation laser system based on an ultra-low-expansion (ULE) glass spacer for a transportable strontium lattice clock \cite{kol17} with an instability of $\mathrm{mod}\,\sigma_{\rm y} = 3 \times 10^{-16}$.

\section{Resonator design and seismic sensitivity}
The optical resonator comprises fused silica substrates (plane and $1$~m radius of curvature) with dielectric Ta$_2$O$_5$/SiO$_2$ high reflectivity coatings optically contacted to an ULE glass spacer ($L=$~120~mm, $D=$~60.8~mm, central bore diameter: 10~mm). 
Fused silica substrates cause lower thermal noise then ULE glass substrates.
To compensate the influence of the higher thermal expansion coefficient of the fused silica substrates, additional ULE glass rings are contacted to their backsides \cite{leg10}.
A thermal noise level of $\mathrm{mod}\,\sigma_{\rm y} = 2.8\times 10^{-16}$ for 698~nm light was calculated for this combination using finite-element modelling (FEM) for the spacer and mirror substrates and analytic formulas for the coatings \cite{kes12}.

\begin{figure}[b!]
	\centerline{\includegraphics[width=.9\columnwidth]{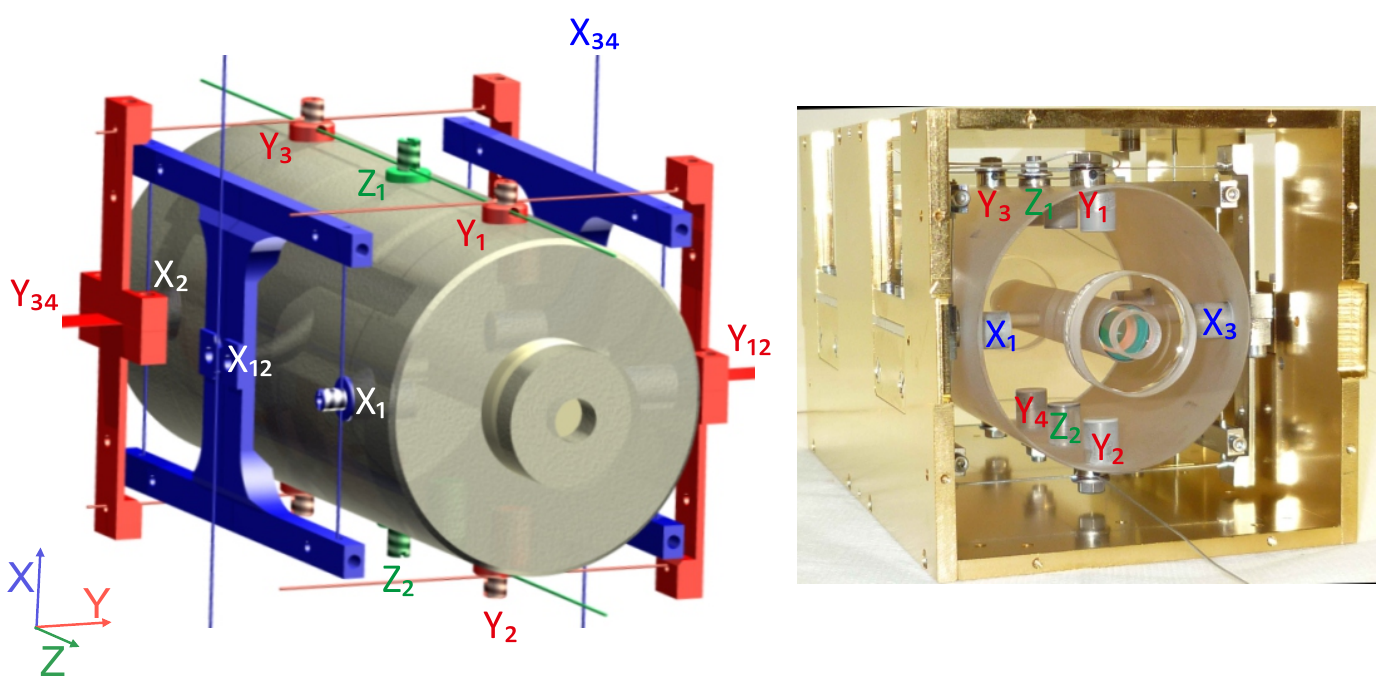}}
	\caption{Left: sketch of the transportable ULE glass cavity. 
		Ten Invar pins are glued into holes to the side of the cavity.
		Flexible wires and stainless steel tapes connect pins, mounting bars and heat shield. 
		For each orientation the pins, wires and bars are color-coded.
		Each color-coded set of mounting elements restricts one rotational and one translational degree of freedom.
		To improve the visibility of the $x$-mounting structure, the red Y-bar in front of the sketch is removed. 
		Right: cavity mounted in its gold plated aluminum heat shield.
		The pins are labelled according to the corresponding sketch.}
	\label{fig:mount}
\end{figure}

The mounting (see Fig.~\ref{fig:mount}) is based on the idea to separate the mounting forces along the three Cartesian directions ($x$, $y$, $z$) \cite{ste11b}.
The corresponding mounting points are placed in the symmetry planes of the cavity ($yz$, $xz$, $xy$).
This configuration minimizes sensitivity of $L$ to accelerations perpendicular to the symmetry planes:
Reversing accelerations, on one side, one would expect the length variation $\Delta L$ to change sign.
On the other side, reversing the accelerations also corresponds to a reflection with respect to the symmetry plane keeping the sign of $\Delta L$.
Consequently, $\Delta L=0$ to first order.
As only forces perpendicular to the symmetry planes act on the mounting points, accelerations do not change the optical length in the central symmetry axis.

Typically, the optical axis defined by the mirror geometry deviates from the symmetry axis of the spacer due to tolerances.
Hence, a tilt of the mirrors will result in a length change of the optical axis \cite{ama13}.
To avoid such tilt, the resonator is mounted at four points equivalent to the Airy points in $x$- and $y$-direction respectively.
With FEM-simulation, the positions were determined to $d = 19.4$~mm from the spacer end faces.
Then the mirrors stay parallel even under acceleration induced bending, avoiding length changes to first order.

This mounting scheme results in ten mounting points.
To hold the cavity body in space, all six degrees of freedom (DOF), three rotations $R_{\rm x}$, $R_{\rm y}$, $R_{\rm z}$ and three translations $T_{\rm x}$, $T_{\rm y}$, $T_{\rm z}$, must be fixed.
To avoid an overdetermined mounting where constraining forces would deform the cavity, the four mounting points in $x$- and $y$-direction are effectively combined by fixing the sum of their individual translation by a lever e.g. $(T_{\rm x}(X_1)+T_{\rm x}(X_2))/2 = T_{\rm x}(X_{12}) = 0$ while relative motion $T_{\rm x}(X_1) -T_{\rm x}(X_2) \neq 0$ is allowed (see H-shaped blue bars Fig.~\ref{fig:mount}, which also constrain rotation $R_{\rm z}$).
Table~\ref{tab:DOF} summarizes the reduction of constrains from ten to six.
\begin{table} [b!]
	\begin{center}
		\begin{tabular*}{10cm}{lll}
			\hline 
			direction of  	& ~~~restrictions 	&  	fixed DOF	\\ 
			acceleration	&																		&						\\
			\hline 
			x 				&	$\left.
			\begin{array}{l}
			\text{combined with blue bars:} \\
			\left[T_{\rm x}(X_1)+T_{\rm x}(X_2)\right]/2 = T_{\rm x}(X_{12})\\
			\left[T_{\rm x}(X_3)+T_{\rm x}(X_4)\right]/2 = T_{\rm x}(X_{34})\\
			\end{array}
			\right\}$
			& 
			$T_{\rm x},\; R_\mathrm{z}$
			\\																				
			\\ 
			y 				&	$\left.
			\begin{array}{l}
			\text{combined with red bars:} \\
			\left[T_{\rm y}(Y_1)+T_{\rm y}(Y_2)\right]/2 = T_{\rm y}(Y_{12})\\
			\left[T_{\rm y}(Y_3)+T_{\rm y}(Y_4)\right]/2 = T_{\rm y}(Y_{34})\\
			\end{array}
			\right\}$
			& 
			$T_{\rm y},\; R_\mathrm{x}$
			\\																				
			\\ 
			
			z 				&	$\left.
			\begin{array}{l}
			\text{direct wire connection:}\\
			T_{\rm z}(Z_1), \; T_{\rm z}(Z_2)\\
			\end{array}
			\right\}$
			&   
			$T_{\rm z}, \; R_\mathrm{y}$
			\\																																															
			\hline \hline
		\end{tabular*}
		\caption{\label{tab:DOF}Summary of merged fixed points to avoid overdetermination of degrees of freedom (DOF).}
	\end{center}
\end{table}
The levers combining the fix points in $x$- and $y$-direction are connected to wires (points $X_{12}$ and $X_{34}$) or metal tapes (points $Y_{12}$ and $Y_{34}$), which are clamped to the inner heat shield using grub screws and small metal clamps.
For reasons of space limitation, the implementation of $x$ and $y$ mounting bars is realized differently, which has no consequence for the reduction of mounting boundary conditions.
For example, the rotations $R_{\rm x}$, $R_{\rm y}$ ($R_{\rm y}$, $R_{\rm z}$) are still allowed by $X_{12}$ and $X_{34}$ ($Y_{12}$ and $Y_{34}$).
For the $z$-direction the green wires attached to the points $Z_1$ and $Z_2$ are connected to the inner heat shield.

In our setup, flexible stainless steel stranded wires are used for mounting, because they allow a fixed mounting along the wire but do transfer smaller forces perpendicular to the wire directions compared to solid wires.
Every fixed point is realized through an Invar pin glued into blind holes ($d = 8$~mm $l = 10$~mm) at the lateral cylinder surface. 
The wire is attached with washers and nuts to the threaded Invar pin heads.
To prevent slacking of the wires, pre-stress on each wire was applied before attaching it to the bar or the heat shield.
This procedure also ensures uniform tension on all wires, which is moderately important for low acceleration sensitivity, as remaining elastic forces transverse to the wires become equal.
Separate tests have yielded a strength of more than $230$~N for every wire connection, where the clamping points were the weakest parts.
Since every DOF of the cavity ($m = 0.83$~kg) is fixed by at least two wires or two flexible tapes a maximum allowable acceleration of $55$~g can be expected.

The seismic sensitivity was measured by sequentially exciting periodic accelerations with a drive frequency of 10~Hz approximately along each Cartesian direction to the cavity and by observing the variation of the beat note frequency obtained with a second reference laser.
In parallel, the actual accelerations in each direction were measured to remove effects of crosstalk between the directions and to disentangle the sensitivity coefficients.
Acoustic excitation of the setup indicates a first resonance at 150~Hz likely due to wire resonances.
Without any further optimization, a vibration sensitivity of 
$\kappa_{\rm x} = 0.7\times 10^{-10}~$/g, 
$\kappa_{\rm y} = 2.3\times 10^{-10}~$/g and
$\kappa_{\rm z} = 12.3\times 10^{-10}~$/g is achieved.
The measured values along the $x$ and $y$ directions are competitive with other transportable systems with $10$~cm cavity length \cite{che14a, swi16}.
The higher value along $z$ most likely originates from inclined wires between the X-pins and the H-bar (colored in blue Fig.~\ref{fig:mount}). 
Through a small tilt of the wires, acceleration along the x-axis causes a direct axial compression or expansion.
No further optimization was necessary due to the overall good laser stability (see section~4).

From the measured sensitivities, a vibration noise floor of $0.2~\mu$g should not be exceeded in order to reach the thermal noise floor at low averaging times.
Thus, the cavity is placed on a vibration isolation table (TS-150, The Table Stable Ltd.), which reduces vibrations by more than 20~dB for $>5$~Hz.

\section{Thermal design}
The length stability for long averaging times is limited by temperature fluctuations and ageing of the glass. 
The use of low expansion material like ULE-glass and a multi-stage temperature stabilization reduces the impact of external temperature fluctuations on the cavity instability. 

\begin{figure}[t]
	\centerline{\includegraphics[width=.9\columnwidth]{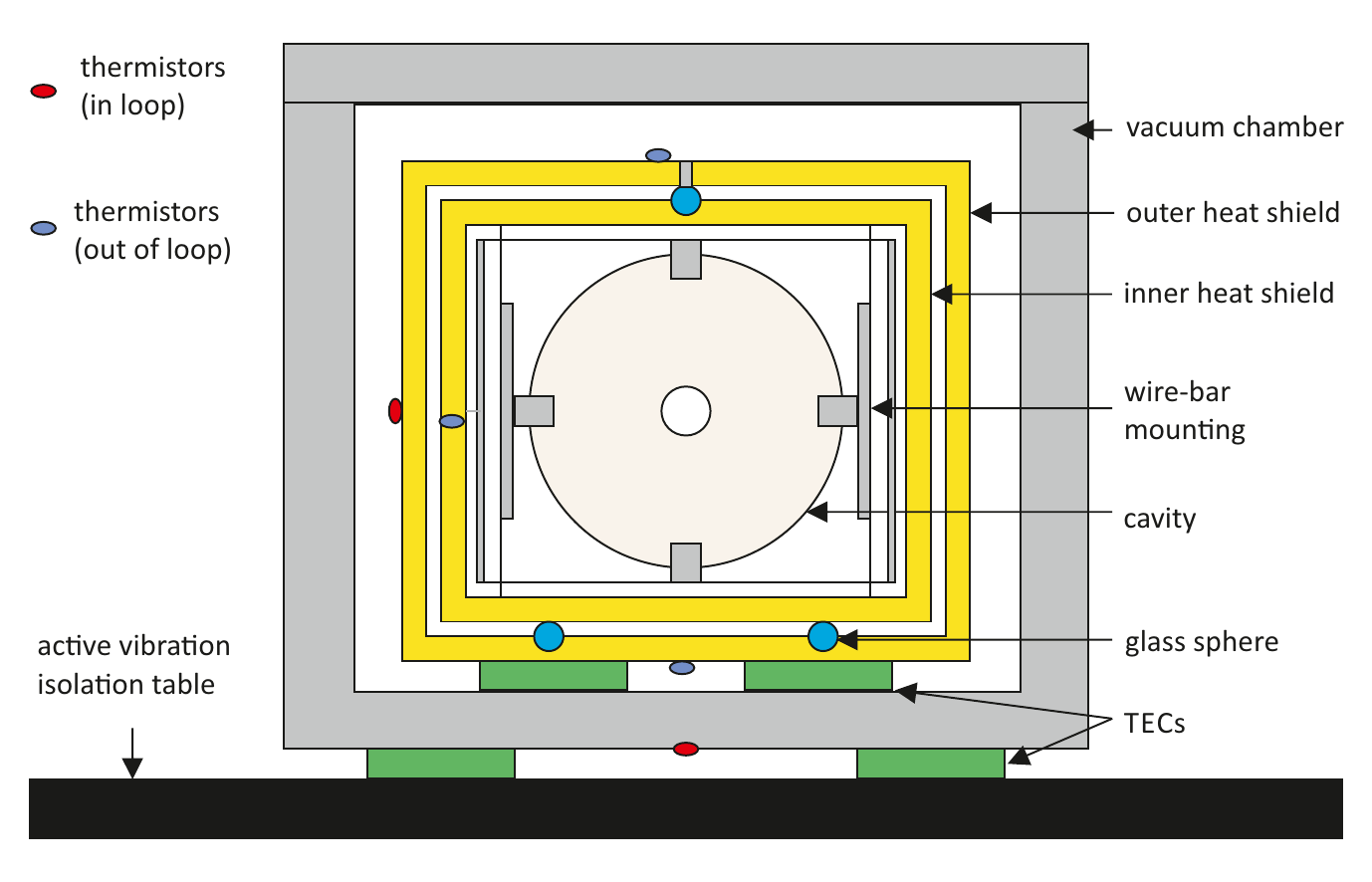}}
	\caption{Cross section of the vacuum chamber, the thermal isolation and the cavity.
		The vibration isolation table serves as a heat sink for the TECs.
		In addition to the thermistors for the temperature control further sensors exist for monitoring.}
	\label{fig:chamber:heatshild}
\end{figure}

During field studies, higher temperature fluctuations than under well controlled laboratory environment arise.
To overcome the influence of such temperature changes, the cavity with its passive heat shield (Fig.~\ref{fig:mount} right) is placed in an actively temperature stabilized outer heat shield, which is itself placed in a temperature stabilized vacuum chamber made out of aluminum (Fig.~\ref{fig:chamber:heatshild}).
The bottom of the outer heat shield is screwed to the bottom of the vacuum chamber to hold it in place.
The inner heat shield is held by four glass spheres in grooves and holes (three at the bottom, one at the top) and a screw pressing on the upper sphere. 
Great effort was taken to realize a compact ($l = 27$~cm; $w = 19$~cm $h = 20$~cm), low mass ($8$~kg) and robust system.
A 3~l/s ion-getter pump ensures a pressure of $3\times10^{-8}~$mbar.
This is low enough to suppress the influence of pressure fluctuation $\Delta p$ on the optical length $L_{\mathrm{opt}} = L~n(p)$ through the refraction index $n$, which depends on pressure.

\begin{figure}[hb!]
	\centerline {\includegraphics[width=.9\columnwidth]{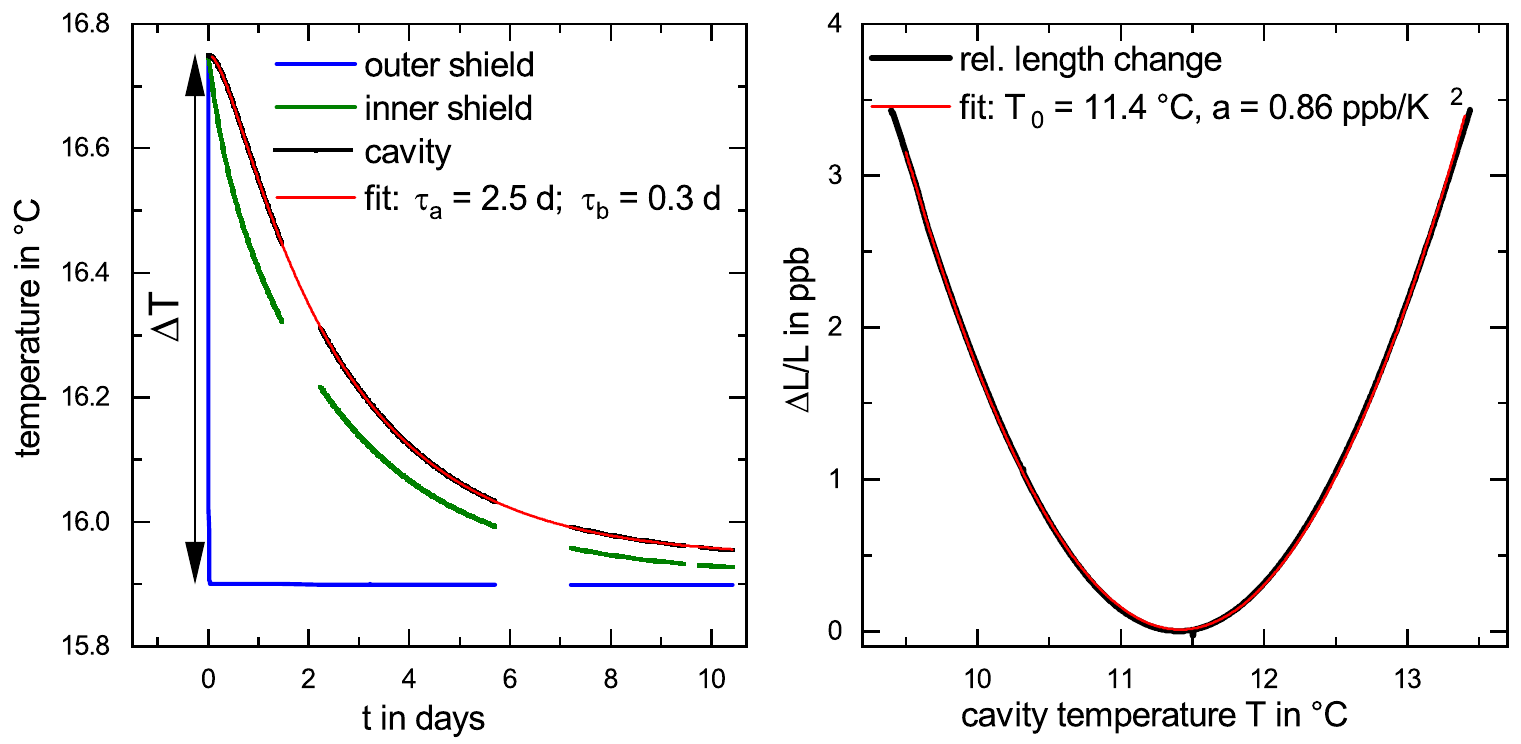}}
	\caption{Left: Thermal step response of the cavity to a rapid change of the temperature $\Delta T$ of the active heat shield (blue). Green lines temperature of the inner shield, black reveals the cavity temperature and in red fit to estimate the time constants $\tau_a$ and $\tau_b$ $\Delta T_{\rm cav} = \Delta T\left(\tau_a/(\tau_b-\tau_a)e^{-t/\tau_a}-\tau_b/(\tau_b-\tau_a)e^{-t/\tau_b}\right)$.
	Right: Measured relative thermal expansion $\Delta L/L$ of the cavity (black).
	The red curve indicates a quadratic fit, $\Delta L/L = a/2\cdot(T-T_0)^2$, to evaluate the CTE.}
	\label{fig:cte}
\end{figure}

The vacuum chamber is temperature stabilized to 20~$^\circ$C using four thermo-electric coolers (TECs) and covered with 2~cm thick Styrofoam.
To protect the system against air flow and acoustic noise, the chamber is placed in a wooden box covered with acoustic insulation foam.

The two heat shields between vacuum chamber and cavity consist of 4~mm thick gold plated aluminum sheets and are separated by four glass spheres to ensure robust mounting and high thermal isolation. 
The outer shield is actively temperature stabilized to the zero crossing temperature $T_0$ of the cavity's coefficient of thermal expansion (CTE) using three TECs and one thermistor placed at the side of the inner heat shield to measure the average shield temperature independent of vertical gradients \cite{hae15b}.
The stationary heat flux resulting from heat radiation and heat conduction via the mounting screws between the chamber and the shield is calculated to be 180~mW for $10$~K temperature difference.
Thermistors are used to control and monitor the temperature of the system.
A typical temperature stability better than 0.2~mK over one day was achieved in a laboratory environment with typical daily temperature fluctuations of 0.1 K.
NBK7-windows in the vacuum chamber and the two heat shields provide optical access to the cavity.

To determine the CTE zero crossing temperature, we have slowly changed the cavity temperature and measured the thermal expansion by recording the beat note of a cavity stabilized laser with an additional reference laser.
The CTE zero crossing temperature is approximately at $11.4~^\circ$C (see Fig \ref{fig:cte} right), as determined by the thermistor at the inner heat shield, which has practically the same temperature as the cavity.
No additional sensor is placed on the cavity in order to avoid a direct seismic and thermal coupling.

To characterize the time constants of our setup, we applied a temperature step from $16.75~^\circ$C to $15.9~^\circ$C to the active heat shield and measured the temperature response of the inner heat shield and the cavity as inferred from its length change via the previously determined CTE.
In Fig.~\ref{fig:cte} left, the thermal step response of each shield and the cavity is displayed.
A fit of the measured data reveals the thermal time constants $\tau_a = 2.5$~days and $\tau_b = 0.29$~days between the outer heat shield and the cavity.
On the one hand, the time constants should be long enough to ensure a good thermal isolation.
On the other hand, the CTE zero crossing temperature must be reached in a reasonable time scale after transportation. 
The usual setup time of the transportable clock system takes about 4-5 days \cite{kol17}.
At constant temperature, the drift of the cavity is approximately 0.1~Hz/s, which can easily be corrected by the clock frequency control.

\section{Optical setup and cavity performance}

\begin{figure}[hbt!]
	\centerline{\includegraphics[width=0.9\columnwidth]{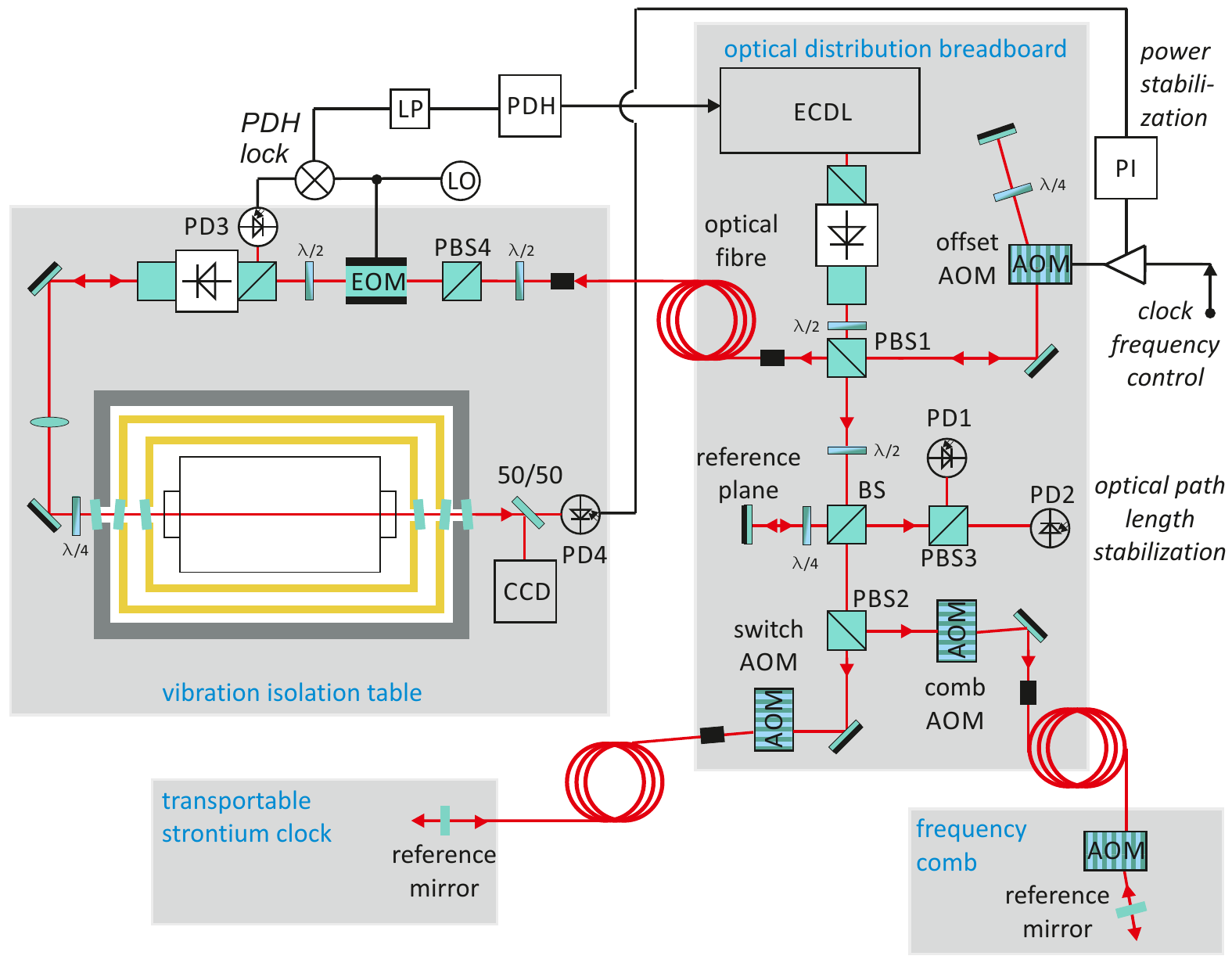}}
	\caption{Experimental setup of the interrogation laser system. 
		Red lines indicate optical signals, black lines: electronic signals, $\lambda/4$: quarter wave plate, $\lambda/2$: half wave plate, PBS: polarizing beam splitter, LP: low pass, CCD: camera, PI: proportional-integral servo, PDH: Pound-Drever-Hall servo.}
	\label{fig:opticalsetup}
\end{figure}

In our setup (Fig.~\ref{fig:opticalsetup}), we use a 698~nm extended cavity diode laser (DLPro, Toptica).
Most of the laser power is transmitted by PBS1 and sent to the clock and the frequency comb.
In this path, the light is split at BS to a reference mirror, which is used as a common reference plane for the heterodyne optical fiber noise cancellation \cite{ma94, sto04} to the atoms and to the comb.
PD1 and PD2 detect the phase fluctuation caused by optical path length variations to the atoms and to the comb, respectively.
Since the light is only sent to the physics package during the atomic excitation time, a pulsed noise cancellation scheme is implemented \cite{fal12}. 

The light reflected at a polarizing beam splitter (PBS1) is sent through an acousto-optical modulator (offset-AOM) in double-pass configuration to allow for a variable frequency offset between cavity and atoms, and then through a polarization maintaining fibre to the cavity setup.
A free space electro-optical modulator (EOM) is used to generate the PDH side bands at $f_{\rm eom}=24$~MHz.
The light reflected back by the cavity is coupled out at an isolator and sent to a photo diode (PD3).
The signal of PD3 is mixed with the radio frequency of the local oscillator to generate the PDH error signal.
Fast frequency variations are corrected via the diode current, slow ones by a piezo which is changing the extended cavity length.

\begin{figure}[bt!]
	\centerline{\includegraphics[width=1\columnwidth]{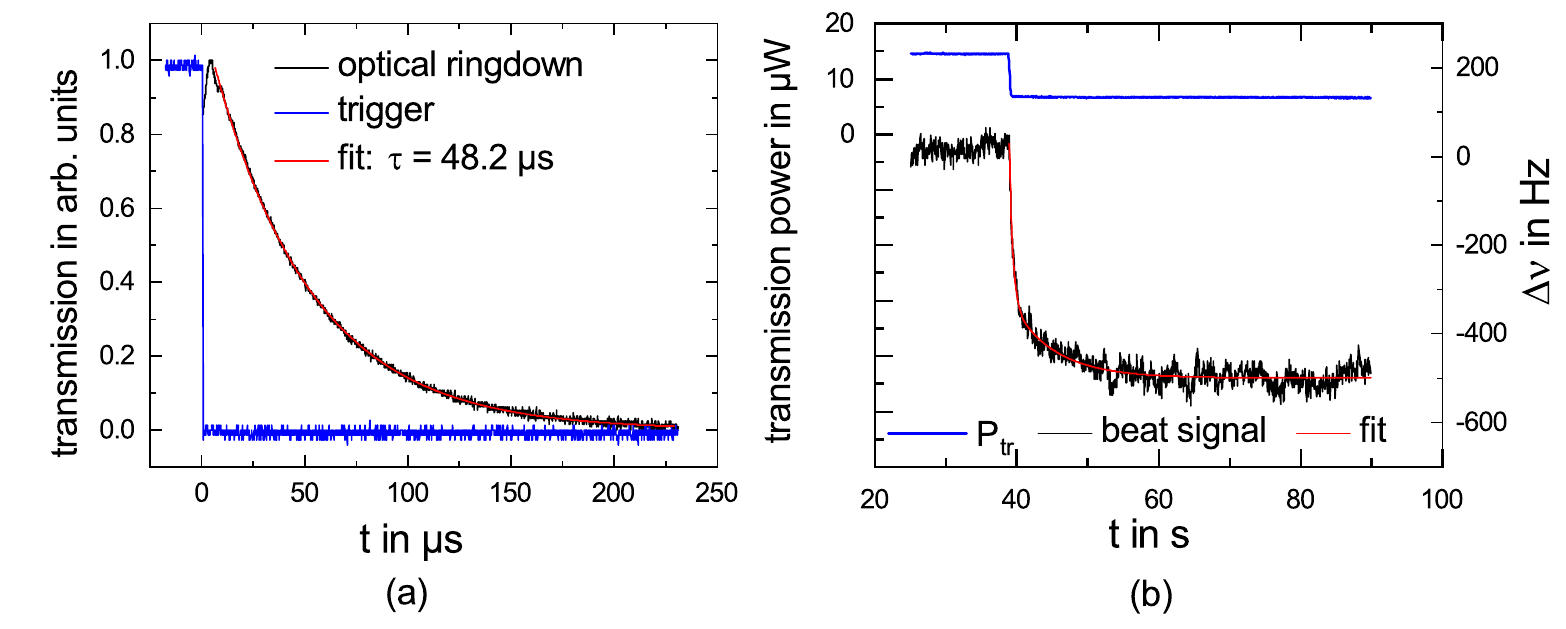}}
	\caption{ (a) Finesse measurement: exponential decay (black) of the transmitted light after switching off with a trigger signal (blue), exponential fit $P(t)=P_0e^{-t/\tau}$ to determine the time constant $\tau$ (red) (b) Measurement of power dependence: the laser frequency shift (black) while changing the transmitted power (blue), exponential fit: $\Delta \nu(t) = \nu_0 + c_1 e^{-t/\tau_1} + c_2 e^{-t/\tau_2}$ (red)}
	\label{fig:ringdown}
\end{figure}

The cavity finesse is measured by the optical ring-down method.
The light is coupled into the cavity and switched off rapidly via the offset-AOM.
From the exponential decay time $\tau$ of the transmitted light the finesse was determined to $F= 460~000$ (Fig.~\ref{fig:ringdown} a).

In the mirror coatings a fraction of the laser power is absorbed and causes thermal deformation of the coatings and substrates, which results in frequency changes.  
In Fig.~\ref{fig:ringdown} b, the frequency response to a power change is displayed.
We observe two different decay time constants $\tau_1 = 0.5$~s and $\tau_2 = 6.1$~s originating from the different thermal time constants of the mirror coating and the substrate \cite{swi16, far12}. 
For long times $t>\tau_2$, the sensitivity is $65~\rm{Hz}/\mu\rm{W}$.
To avoid corresponding frequency fluctuations, the transmitted laser power (PD4) is stabilized using the offset-AOM.
This allows us to suppress the contribution of power fluctuations to laser frequency noise to a level below $\mathrm{mod}\,\sigma_{\rm y}= 1\times 10^{-16}$ for averaging times between 1~s to 100~s.
Without power stabilization, an instability of $\mathrm{mod}\,\sigma_{\rm y}= 2\times 10^{-15}$ at one second was observed.

\begin{figure}[h!]
	\centerline{\includegraphics[width=.7\columnwidth]{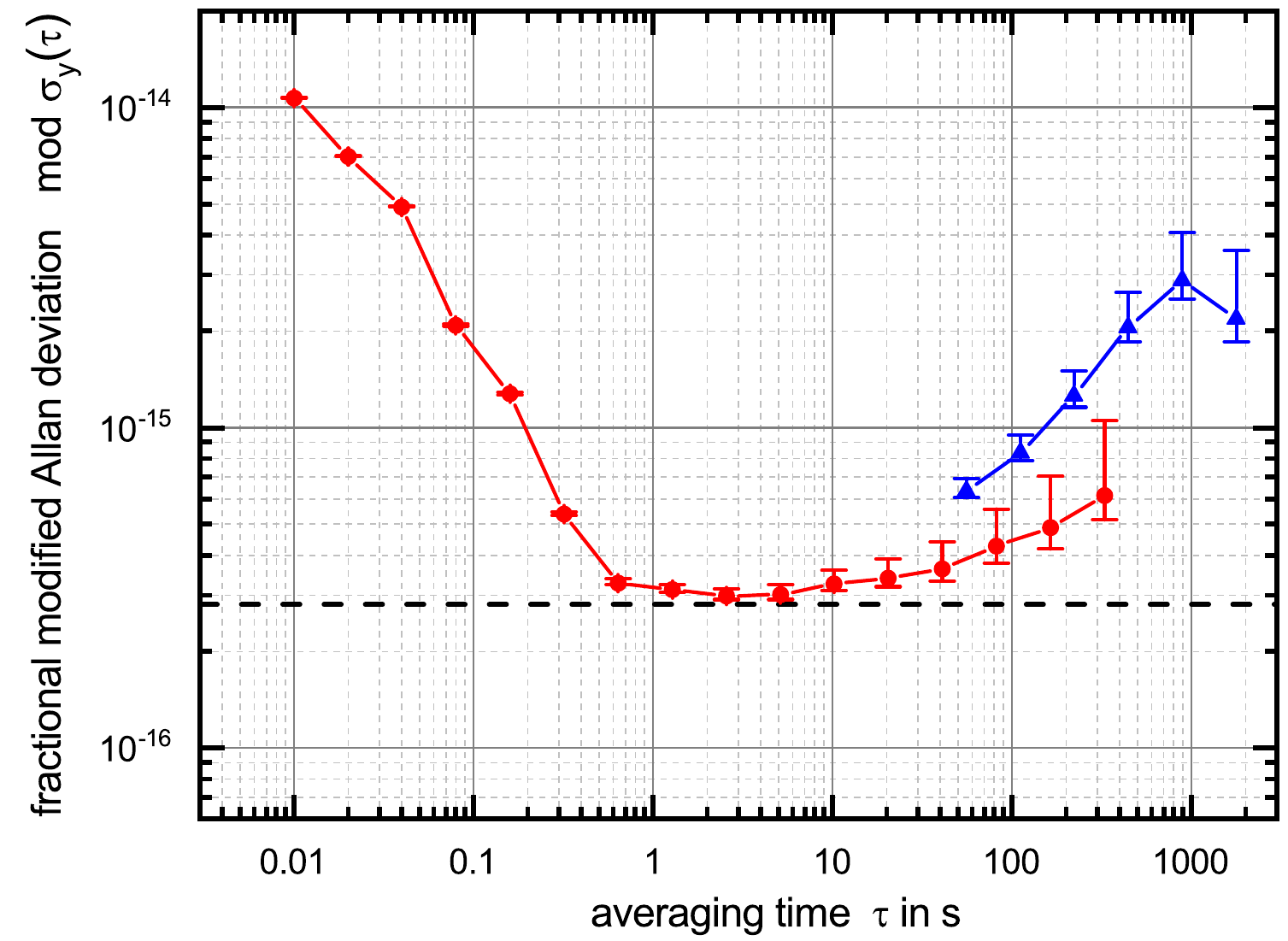}}
	\caption{Measured frequency instability of the interrogation laser compared to the stationary clock laser system (red) and long term stability during field study (blue).}
	\label{fig:allan}
\end{figure}

An EOM generates not only phase modulation, but also residual amplitude modulation (RAM).
This leads to additional fluctuations in the PDH error signal that degrade the laser stability \cite{zha14}.
The home-made EOM consists of a Brewster-cut lithium niobate crystal, which is electrically driven by a resonant RF-transformer.
We found that the laser instability caused by the RAM of this EOM is below $\mathrm{mod}\,\sigma_{\rm y}= 5\times 10^{-17}$ for averaging times between 1~s and 100~s.

We have evaluated the laser instability by using the stationary strontium clock laser system \cite{hae15a}, which in turn is locked to a silicon laser system \cite{mat17a} with an instability of $\mathrm{mod}\,\sigma_{\rm y}=4\times 10^{-17}$. 
The beat note frequency between the stationary and the transportable systems was measured with a dead time free frequency counter.
We observe occasional frequency jumps by about 5~Hz a few times a day.
The origin of the jumps is not clear, but it is most likely that mechanical relaxation of the strained stainless steel wires is responsible.
Those could be replaced by single steel wires to overcome this effect.
A linear drift of 87~mHz/s was subtracted and the modified Allan deviation $\mathrm{mod}\,\sigma_{\rm y}$ was calculated, see Fig.~\ref{fig:allan}.  
It shows an instability of approximately $\mathrm{mod}\,\sigma_{\rm y}= 3\times 10^{-16}$ between 0.5~s and 10~s, which matches the calculated thermal noise floor of $\mathrm{mod}\,\sigma_{\rm y}= 2.8\times 10^{-16}$.

\section{Field study}

In three field studies in Modane/Torino (France/Italy, 2016) \cite{gro18a}, Paris (France, 2017) and Munich (Germany, 2018), the cavity has proven its robustness.
Before, the transportable clock \cite{kol17} was interrogated with this clock laser system placed in a laboratory environment.
An instability of $\sigma_{\rm y}=1.4\times10^{-15}/\sqrt{\tau/\mathrm{s}}$ was determined.
During the field studies, we observed a higher clock instability of  $\sigma_{\rm y}=2 -3\times10^{-15}/\sqrt{\tau/\mathrm{s}}$ than in the well controlled laboratory environment.
It is most likely that the clock laser stability was reduced due to higher seismic noise and temperature fluctuations.
The long term stability ($>100$~s) of our clock laser is determined by  the clock frequency control signal applied to the offset-AOM (Fig.~\ref{fig:allan} blue).
We observe a two times higher instability at an averaging time of 100~s, compared to the laboratory stability.
Especially, the system was degraded at 1~s averaging time by the not noise-compensated fibre between the distributions breadboard and the cavity. 
This impact was reduced by implementing a shorter and passively thermally isolated fibre.
The cavity setup itself is very robust against transportation shocks.
No optical adjustment was needed to operate the cavity after transport.
The laser system is setup within one day after transport and is ready for operation after approximately two days, when the temperature of the cavity has settled.

\section{Conclusion and Outlook}
We presented a compact robust and highly frequency stable interrogation laser system for the transportable strontium lattice clock.
The system is designed to be compact and low-mass while having very small sensitivity to vibration noise ($0.7\times 10^{-10}~$/g to $12.3\times 10^{-10}~$/g).
To our knowledge, this is the first mounting scheme for a transportable system that at the same time realizes independent mounting and optimization of the support points for all directions and avoids overdetermination of the degrees of freedom.
With a measured instability of $\mathrm{mod}\,\sigma_{\rm y}=3\times 10^{-16}$, this system has reached, to the best of our knowledge, the lowest instability for a transportable clock laser. 
To reduce the thermal noise, crystalline coatings \cite{col13} with a low mechanical loss (only available for near infrared) can be implemented.
Those mirrors would boost the laser instability down to $\mathrm{mod}\,\sigma_{\rm y}=1.4\times 10^{-16}$.
In parallel, the vibration sensitivity must be reduced by more careful orientation of the mounting wires and possibly implementing an additional feed forward correction \cite{tho10}.
In addition, temperature fluctuations can be further reduced by actively stabilizing the temperature inside the acoustic insulation box.
To use crystalline mirrors, operation of the cavity at 1397~nm and frequency doubling to 698~nm is necessary. 
Additional noise of the doubler has already been tested to an instability level of $10^{-19}$ at 1~s \cite{her19}.
As the cavity mounting is designed to withstand acceleration shocks of up to 55~g an application of the cavity in space missions \cite{ori16} seems feasible.

\section*{Fundings}
Funded by the Deutsche Forschungsgemeinschaft (DFG, German Research Foundation) under Germany's Excellence Strategy -- EXC-2123 QuantumFrontiers -- 390837967 and CRC 1128 geoQ within project A03.

\section*{Acknowledgement}
We thank Jacopo Grotti and Silvio B. Koller for operating the clock during the field studies and for providing the offset-AOM data.

\section*{Disclosures}
US: Physikalisch-Technische Bundesanstalt (P)

\bibliography{sample}

\end{document}